 \documentclass[preprint,review,12pt,sort&compress]{elsarticle}
\usepackage[margin=1.18 in]{geometry}
\usepackage{multirow}
\usepackage{adjustbox}
\usepackage{graphicx}
\usepackage{gensymb}
\usepackage{xcolor}
\usepackage{dirtytalk}
\usepackage[title]{appendix}
\definecolor{UW}{RGB}{64, 38, 96}
\usepackage{amssymb}
\usepackage{amsmath}
\usepackage{breqn}
\usepackage{textcomp}
\newcommand{\B}{Ba\v zant}  

\newcommand{\bc}{\begin{center}}
\newcommand{\ec}{\end{center}}
\newcommand{\bfr}{\begin{flushright}}
\newcommand{\efr}{\end{flushright}}
\newcommand{\ii}{\item}
\newcommand{\be}{\begin{enumerate}}
\newcommand{\ee}{\end{enumerate}}
\newcommand{\bi}{\begin{itemize}}
\newcommand{\ei}{\end{itemize}}
\newcommand{\bd}{\begin{description}}
\newcommand{\ed}{\end{description}}
\newcommand{\beq}{\begin{equation}}
\newcommand{\eeq}{\end{equation}}
\newcommand{\bea}{\begin{eqnarray}}
\newcommand{\eea}{\end{eqnarray}}

\newcommand{\bfi}{\begin{figure}}
\newcommand{\efi}{\end{figure}}
\newcommand{\bay}{\begin{array}{l}}
\newcommand{\eay}{\end{array}}

\journal{Composite Part A: Manufacturing}

\begin{document}


\begin{titlepage}
\clearpage\thispagestyle{empty}

\noindent
\hrulefill

\begin{figure}[h!]
\centering
\includegraphics[width=1.5 in]{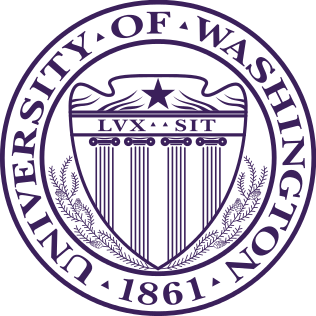}
\end{figure}

\begin{center}
{\color{UW}{{\bf A\&A Program in Structures} \\ [0.1in]
William E. Boeing Department of Aeronautics and Astronautics \\ [0.1in]
University of Washington \\ [0.1in]
Seattle, Washington 98195, USA}}
\end{center} 

\hrulefill \\ \vskip 2mm
\vskip 0.5in

\begin{center}
{\large {\bf Effect of the Thickness on the Fracturing Behavior of Discontinuous Fiber Composite Structures}}\\[0.5in]
{\large {\sc Seunghyun Ko, James Davey, Sam Douglass, Jinkyu Yang, Mark E. Tuttle, Marco Salviato}}\\[0.75in]
{\sf \bf INTERNAL REPORT No. 19-03/02E}\\[0.75in]
\end{center}

\noindent {\footnotesize {{\em Submitted to Composite Part A: Manufacturing \hfill March 2019} }}
\end{titlepage}

\newpage

\begin{frontmatter}



\title{Effect of the Thickness on the Fracturing Behavior of Discontinuous Fiber Composite Structures}

\author[label1]{Seunghyun Ko}
\author[label1]{James Davey}
\author[label1]{Sam Douglass}
\author[label1]{Jinkyu Yang}
\author[label2]{Mark E. Tuttle}
\author[label1]{Marco Salviato \corref{cor1}}
 \address[label1]{William E. Boeing Department of Aeronautics and Astronautics, University of Washington, Seattle, WA 98195, USA}
 \address[label2]{Department of Mechanical Engineering, University of Washington, Seattle, WA 98195, USA}

\cortext[cor1]{Corresponding Author, \ead{salviato@aa.washington.edu}}

\begin{abstract}
\linespread{1}\selectfont
In this study, we investigate experimentally and numerically the mode I intra-laminar fracture and size effect of Discontinuous Fiber Composites (DFCs) as a function of the structure thicknesses. 

By testing geometrically-scaled Single Edge Notch Tension (SENT) specimens a notable structure size effect on the nominal strength of DFCs is identified. As the specimen size increases, the nominal strength decreases. For small specimens, we find a limited size effect with enhanced pseudo-ductility and a strong divergence from Linear Elastic Fracture Mechanics (LEFM). For sufficiently large specimen sizes, the scaling of the nominal strength follows closely LEFM with a strong brittle failure. As the thickness increases, the size effect decreases.

We identify the fracture energy and the effective size of the fracture process zone as a function of the thickness of the structure. To do so, we integrate equivalent fracture mechanics and stochastic finite element modeling. Experimentally, we collect the nominal strength of geometrically-scaled Single Edge Notch Tension (SENT) specimens. The numerical stochastic model captures the complex, inhomogeneous mesostructure of DFCs by explicitly generating the platelets. From the integrated analysis, it is found that the fracture energy depends significantly on the structure thickness. It is shown to increase gradually up to $2$ mm and saturates after $3$ mm to a value of $57.77$ N/mm, which is $4.81$ times larger than a typical aluminum alloy. 



\end{abstract}


\end{frontmatter}




\section{Introduction}

Discontinuous Fiber Composites (DFCs) made of chopped fiber platelets offer unique advantages over traditional unidirectional composites. The possibility of manufacturing DFC components by compression molding thanks to the outstanding formability, makes DFCs a highly viable option even for very complex geometries \cite{Boeing, ForgedComposite, Tencate}. These characteristics have attracted significant interest from the scientific and industrial communities for the use of DFCs in applications that have been typical of light alloys, such as secondary structural components for aerospace \cite{Boeing,Hexcel,Tencate}, body frames of terrestrial vehicles \cite{ForgedComposite, Hexcel, Toyota, Quantum}, composite brackets, suspension arms and interiors \cite{ForgedComposite} and crash absorbers \cite{Quantum, Hexcel, ForgedComposite}. 

Another interesting feature of DFCs is their mechanical behavior compared to more traditional composites. Recent studies have shown that DFCs tend to exhibit a more pseudo-ductile behavior \cite{Io2006, Fera2009a, Fera2009b, Wan2016, Shin2016, Sel2015} and a significantly lower notch sensitivity \cite{KoArxiv, KoASC, KoSampe} compared to laminated composites. At the same time, the fracture energy dissipated by DFCs has been shown to be noticeably larger than aluminum alloys and typical quasi-isotropic laminates, the main competing materials in the aerospace industry \cite{KoArxiv}. 

The foregoing interesting mechanical properties, however, have been shown to be significantly dependent on the material mesostructure and the structure size and geometry. In \cite{Fera2009a}, Feraboli \textit{et al}. conducted an extensive experimental campaign which showed that, for the range of plate thickness and platelet sizes investigated, the tensile and compressive strength increase with the platelet length whereas the effective elastic moduli seem to be almost unaffected. Combining experiments and two-dimensional Finite Element modeling, Selezneva \textit{et al}. \cite{Sel2015, Sel2017} found that the measured strengths and their Coefficient of Variation (CoV) are strongly related to the platelet size and the thickness of the plates. The longer platelets provide higher strength with increased CoV whereas the thicker plates increase the average strength and feature a lower variability. Similar tests were conducted by Wan and Takahashi \cite{Wan2016} who investigated the effects of the molding pressure. They concluded that a higher molding pressure increases the strength in both tension and compression whereas no significant differences were found on the effective elastic moduli. Nilakantan and Nutt \cite{Nila2018} further extended the study on the effects of the processing conditions on the mechanical properties of DFCs. They investigated different manufacturing methods such as open, closed mold, and vacuum-bag-only. They found that the manufacturing conditions affect the mechanical performance of DFCs in a substantially different manner compared to continuous fiber composites. For instance, they found that DFCs show little correlation between the void contents and the tensile strength. Leveraging a three-dimensional mesoscale model and experimental tests, Kravchenko \textit{et al}. \cite{Pipes2018, Pipes2019} investigated notch-free DFC structures with aligned and staggered platelets. Using controlled platelets orientations, they found that DFCs with longer and thinner platelets have higher strengths. They also showed that the mechanisms of failure transition from delamination-dominated fracture to the fracture by platelet failure as its length increases. This phenomenon, similar to the one reported in other staggered materials systems such as nacre \cite{Barth2014, Gao2003}, causes the strength to increase almost linearly with the platelet size until reaching an asymptotic value. This asymptotic value depends on the strength of the platelet and its morphology and the mechanical properties of the polymer matrix. 

Ko \textit{et al}. \cite{KoArxiv} showed that not only the platelet morphology but also the structure size and geometry affect the fracturing behavior. Through a size effect test campaign on geometrically-scaled Single Edge Notch Tension (SENT) specimens made with platelets of different sizes, they were able to show that the fracturing behavior changes from pseudo-ductile to brittle with increasing specimen sizes. They also proposed a framework combining equivalent fracture mechanics and stochastic finite elements to estimate the fracture energy of the material. They found that this energy increases almost linearly with increasing platelet sizes. An aspect that was not investigated, however, was the effect of the thickness of the structure on the overall fracturing behavior and its scaling.

In order to shed more light on this critical aspect, this study presents an investigation of the intra-laminar fracture and the size effect in DFCs for four distinct structure thicknesses. The size effect on the structural strength of geometrically-scaled SENT specimens is characterized for each thickness. To capture the size effect and characterize the fracture energy, $G_f$, and the effective length of the fracture process zone, $c_f$, for each thickness, the approach combining equivalent fracture mechanics and stochastic finite element modeling proposed by Ko \textit{et al}. \cite{KoArxiv} is used in this study. The model accounts for the complex random mesostructure of the material by modeling the platelets explicitly. Thanks to this theoretical framework, the mode I fracture energy of DFCs is estimated and it is shown to depend linearly on the thickness until it reaches an asymptotic value of about $58$ N/mm. For all the thicknesses investigated in this work, the fracture energy was much larger than aluminum alloy and quasi-isotropic laminate composites made from the same constituents. This aspect is significantly interesting considering the possible use of DFCs in applications demanding crashworthiness (see e.g. \cite{Matsuo2017}).

  

\section{Material preparation and test description}

\subsection{Material preparation}
In previous studies \cite{Sal2016a, KoArxiv, Baz1996, Sal2016c, Kir2016}, size effect testing of geometrically-scaled SENT specimens led to an accurate measure of mode I intra-laminar fracture properties of unidirectional laminates, textiles, and DFCs. Therefore, in this study, we followed the experimental procedures proposed by Ko \emph{et al}. \cite{KoArxiv}. As a first step, we manufactured the specimens using a Toray T700G-12K prepreg system and leveraging the in-house manufacturing process reported in \cite{KoSampe}. We control the weight of the plate to manufacture consistent plate thickness.

To investigate the effect of the thickness, four different plate thicknesses were investigated: $4.1$, $3.3$, $2.2$, and $1.1$ mm. These values are chosen considering typical applications of DFCs. For each thickness, four different geometrically-scaled specimen sizes were investigated with all the dimensions being scaled (Fig.~\ref{f-Geometry}), while keeping the platelet size constant ($50 \times 8$ mm). As summarized in Table~\ref{T1}, the ratio between the largest (size-$1$) and the smallest (size-$4$) specimen was $12.3$ to $1$, a ratio considered to be sufficiently large to enable a thorough investigation of the size effect on the nominal strength.

To create a notch, a diamond-coated razor saw of width $0.2$ mm was used. We choose to saw the specimens because it prevents the development of the Fracture Process Zone (FPZ) before the tests \cite{Sal2016a}. For all the tests, the size of the initial notch $a_0$ was geometrically-scaled and equal to $D/5$. After manufacturing the notch, a layer of white paint followed by the black speckles was sprayed to perform the Digital Image Correlation (DIC) measurements.

\subsection{Testing}
We tested the specimens using a servo-hydraulic, closed-loop Instron $5585$H with $200$ kN capacity. To prevent strain rate effects, a $0.2~\%$/min nominal strain rate calculated as $\epsilon_{Nom} = \delta/L$ where $\delta = $ displacement from the Instron machine and $L = $ gauge length was used for all the specimens. The load was recorded with a sampling frequency of $10$ Hz. In addition, we captured the digital images with a sample rate of $1$ Hz using a Nikon D5600 DSLR camera with two lenses: A Nikon AF micro $200$ mm and a Sigma $135$ mm DG HSM. All the digital images were analyzed using GOM Correlate software \cite{GOM}. We calculated the nominal displacement, $u$, by taking the relative average displacement between two horizontal lines across the width placed symmetrically with respect to the notch. To scale this nominal displacement with the specimen size, the distance from the notch was chosen to be $1.2$ times $D$.


\section{Experimental results}
\subsection{Load-displacement curves} 


The load and the nominal displacement curves were analyzed after the fracture tests (Fig.~\ref{f-FDdiagram}a-d). Similar to other quasibrittle materials \cite{Baz1984, Baz1990, Baz1998, Yao2018a, Yao2018b, Yao2018c, Cusatis2018}, a strong size effect can be observed in the load-displacement curves reported in Figs.~\ref{f-FDdiagram}a-d for all the thicknesses. For instance, in Fig.~\ref{f-FDdiagram}a, the size-$1$ (largest) shows an evidently linear behavior up to the peak load followed by sudden failure and snap-back instability \cite{Sal2016b}, a typical phenomenon of a brittle structure. However, as the specimen size decreases, the load-displacement curve becomes increasingly non-linear before the peak load (see size-$4$ curves in insets). This is because, as the structure size decreases, the relative size of the FPZ increases and the effects of the sub-critical damage induced in the FPZ becomes more significant. 

Another aspect to notice is the increased scatter of the strength as the structure thickness decreases. For DFCs with $1.1$ mm thickness (see Fig.~\ref{f-FDdiagram}d), it is difficult to distinguish the failure loads between sizes $2$ and $3$ due to significant scatter in the experimental curves. In contrast, for higher thicknesses, the CoV of the data is significantly lower (see Fig.~\ref{f-FDdiagram}a). This can be qualitatively explained by considering the mesoscale morphology of DFCs as a function of the structure thickness. We will address this point in section 5.1.



\subsection{Fracture surfaces}
Representative intra-laminar fracture surfaces of the tested specimens are plotted in Figs.~\ref{f-FractureSurfaces}a-l. Multiple damage mechanisms can be observed including platelet delaminations, matrix microcracking, fiber pull-outs, and platelet splitting and breakages. Further, the fracture paths and the location of the fracture initiation depend on the structure size and thickness relative to the platelet size. For size-$1$ (see Figs.~\ref{f-FractureSurfaces}a-d), the fracture paths are almost perpendicular to the loading direction. As the structure size decrease (see Figs.~\ref{f-FractureSurfaces}e-l), the paths become more tortuous and random. The location of fracture initiation also depends highly on the size of the structure. A higher probability of fracture away from the notch can be observed for the smaller sizes (see Figs.~\ref{f-FractureSurfaces}k,l) due to the larger size of the FPZ compared to the structure dimensions. In fact, the stress redistribution in the FPZ can mitigate the severity of the notch, thus promoting fracture initiation from other weak spots such as resin rich areas, air pockets, and locations where the platelets are unfavourably oriented with respect to the loading direction. 

In addition to the structure size, the structure thickness also influences the process of fracture initiation and propagation. This is clearly shown in Figs.~\ref{f-FractureSurfaces}d,h,l where the DFC specimens featuring the lowest thickness exhibit pronounced fiber breakages and a relatively \say{clean} fracture path. Moreover, as shown in Fig.~\ref{f-FractureSurfaces}d, the fracture happens away from the notch even for size-$1$. For larger thicknesses, the notch size featured by size-$1$ specimens is large enough to make it trigger the fracturing process. Yet, for the specimens with the lowest thickness, the fracture was still substantiated by other weak spots.

\subsection{Structural size effect on the nominal strength}
We define the nominal strength as the failure load, $P_c$, over the gross cross-section: $\sigma_{Nc}=P_c/Dt$ where $D$ and $t$ are the specimen width and thickness. The experimental results of the average nominal strength are summarized in Table~\ref{T2}. As can be noted, the average strength clearly shows a decreasing trend as the structure size increases for all the thickness investigated in this work. To analyze such scaling effect, the following sections introduce a theoretical framework combining equivalent linear elastic fracture mechanics and stochastic finite element analysis.


\section{Theoretical Framework}
To capture the scaling of the structural strength in DFC structures weakened by stress-free notches, we combine an equivalent fracture mechanics approach featuring a characteristic length-scale accounting for the finite size of the FPZ and stochastic finite element analysis to account for the random distributions of the platelets \cite{KoArxiv, Cusatis2018}. 

\subsection{Analysis of size effect tests by Size Effect Law (SEL) for DFCs}

To consider the FPZ, an equivalent crack length, $a$, is defined as the sum of the initial crack length, $a_0$ (see Fig.~\ref{f-Geometry}) and an effective FPZ size, $c_f$, which is assumed to be a material property:
\begin{equation}\label{eq1}
a = a_0 + c_f
\end{equation}
Based on Linear Elastic Fracture Mechanics (LEFM), the energy release rate, $G$ can be written as a function of the dimensionless crack length, $\alpha = a/D$:
\begin{equation}\label{eq2}
G(\alpha) = \frac{\sigma_N^2 D}{E^*}g(\alpha)
\end{equation}
Here, the nominal stress is $\sigma_{N}=P/Dt$ where $P$ is the load, $D$ is the specimen width, and $t$ is the specimen thickness, $E^*$ is an effective elastic constant, and $g$ is the dimensionless energy release rate which typically accounts for the effects of the geometry on the energy release rate. If the structure is homogeneous, $g$ depends only on the geometry of the structure and takes the same value for geometrically-scaled specimens regardless of the structure size \cite{Baz1996, Sal2016a}. However, DFCs are highly inhomogeneous materials, the inhomogeneity dimensions being comparable with the size of typical DFC structures. Accordingly, different structure sizes may lead to a significantly different material morphology, thus affecting the energy release and making $g$ dependent on the structure size, $D$ and the thickness $t$.

We can rewrite the Eq.~\eqref{eq2} to account for the inhomogeneity of DFCs as follow \cite{Baz1996, Sal2016a}:
\begin{equation}\label{eq3}
G(\alpha, D) = \frac{\sigma_N^2 D}{E^*}g(\alpha, D)
\end{equation}
where the effect of the size $D$ on the dimensionless energy release rate is explicitly introduced. In contrast, $t$ is not introduced in the equation since Eq.~\eqref{eq3} is used here to analyze geometrically-scaled specimens of the same thickness. Accordingly, a different set of dimensionless energy release rate functions, $g\left(\alpha,D\right)$ needs to be calculated for every thickness investigated in this work.

At incipient failure, $G$ must be equal to the fracture energy, $G_f$, which is assumed to be a material property. Substituting Eq.~\eqref{eq1} into Eq.~\eqref{eq3}, we get:
\begin{equation}\label{eq4}
G_f = G(\alpha_0 + c_f/D, D) = \frac{\sigma_{Nc}^2 D}{E^*}g(\alpha_0 + c_f/D, D)
\end{equation}
where $\sigma_{Nc}$ is the nominal strength at failure. We further expand Eq.~\eqref{eq4} using a Taylor series expansion at $\alpha_0$ with a constant $D$. The following equation is derived:
\begin{equation}\label{eq5}
G_f = \frac{\sigma_{Nc}^2 D}{E^*}\Big[g(\alpha_0, D) + \frac{c_f}{D}\frac{\partial g}{\partial \alpha}(\alpha_0, D)\Big]
\end{equation}
Eq.~\eqref{eq5} can be rearranged into the following form:
\begin{equation}\label{eq6}
\sigma_{Nc} = \sqrt[]{\frac{E^* G_f}{Dg\left(\alpha_0, D\right)+c_f g_D\left(\alpha_0, D\right)}}
\end{equation}
where $g_D(\alpha_0, D) = \frac{\partial g}{\partial \alpha}(\alpha_0, D)$. The subscript $D$ corresponds to the partial differentiation with respect to $\alpha$ for a constant structure size $D$. It is noteworthy that Eq.~\eqref{eq6} contains all the necessary components to capture the scaling of the nominal strength of DFC structures. The characteristic FPZ size, $c_f$, enables to capture the transition from quasi-ductile to brittle fracturing behavior with increasing the structure size. Further, the dimensionless energy release rate parameters, $g$ and $g_D$ combine the geometrical effect as well as the inhomogeneous material characteristics. These parameters can be calculated using the stochastic finite element method described in the following section.

\section{Stochastic finite element model}

In order to characterize the fracture energy, $G_f$ and the effective FPZ length, $c_f$, we need to calculate the dimensionless energy release rate parameters, $g(\alpha_0,D)$ and $g_D(\alpha_0,D)$. These two parameters are related to a release of elastic energy caused by the creation of new crack surfaces. Due to the inhomogeneous mesostructure, the release of energy is strongly related to two elements in addition to the geometry of the structure: the platelet constitutive properties, and the orientation distribution of the platelets. Therefore, modeling explicitly the distribution of the platelets is key. More details on the algorithm and its implementation can be found in \cite{KoArxiv} and Appendix \ref{AppxA}.

Using the mesostructure generation algorithm, we first investigate the mesoscale morphology of DFCs with respect to the structure thickness. The distributions of the average platelet orientations through the thickness are considered. We generate a size $1$ specimen ($L=178, D=80$ mm) for $4$ different thicknesses using the same Cumulative Distribution Function (CDF) of in-plane platelet orientations. The range of the platelet orientation is $-90^\circ \leq \theta \leq 90^\circ$. We calculate the average absolute platelet orientations through thickness, $\theta_A$, in each partition. 
Figure~\ref{f-pdf} presents the Probability Density Function (PDF) of $\theta_A$ for each thickness. The bar graphs represent the frequency of $\theta_A$ from the mesostructure generation algorithm whereas the solid line represents a Gaussian probability fitting. As can be noted, the mean $\theta_A$ stays near $\approx45^\circ$ regardless of the structure thickness. However, the standard deviation changes dramatically with respect to the thickness, from $4.01^\circ$ for $t = 4.1$ mm to $8.59^\circ$ for $t = 1.1$ mm. To visualize the differences, we calculate the probability of $\theta_A \geq 55^\circ$, represented as shaded areas in Fig.~\ref{f-pdf}. For $t=4.1$ mm, the probability is $0.56\%$ which is tremendously small comparing to $16.42\%$ for $t=1.1$ mm. This study quantitatively confirms that the probability of having unfavorable platelet orientations increases as the structure thickness decreases if the CDF of in-plane platelet orientations remains consistent. As a result, the average strength of DFCs deviates significantly in lower thickness as we observed from the experiments. It is worth mentioning that the results of this qualitative investigation are related to DFC plates where the distribution mechanism of platelets is not affected by the thickness.
A different scenario can be expected in structures featuring complex geometrical features in which high platelet flow can occur. In such a case, the effect of the thickness of the structure must be investigated via an accurate morphological study.

The completed mesostructure is now transferred to Abaqus/Standard \cite{Abaqus} to find the dimensionless energy release rate functions. See Appendix \ref{AppxB} for the detailed implementation. Due to the stochastic nature of the problem, a total of $10$ different DFC structures for each specimen sizes are generated to find the average $g$ and $g_D$. The process is repeated for all $4$ different thicknesses. It is worth mentioning that additional $10$ structures were generated for the thickness of $1.1$ mm due to its high variations found in $g$ and $g_D$. A summary of all the results is provided in Fig.~\ref{f-gandgprime} whereas the mean and standard deviation for each thickness is given in Table~\ref{T4}. As can be noted, for thicknesses higher than $2.2$ mm, the effect of the structure size in $g$ and $g_D$ is negligible. Especially, for the thickness of $4.1$ mm, there is hardly no dependency in $g$ and $g_D$ with the size of the structure. This is because for sufficiently large thicknesses spatial distribution of platelets becomes almost uniform, and the in-plane behavior becomes closer and closer to the one of an homogeneous medium. As a result, $g$ and $g_D$ do not change for the geometrically-scaled structures. However, the effect of the material inhomogeneity remains strong for the $t = 1.1$ mm. As can be noted from Fig.~\ref{f-gandgprime}d, the $g$ and $g_D$ change with respect to the structure size with large deviations. This is consistent with the observations made in section 3.1 and Fig.~\ref{f-FDdiagram}.
Consequently, we need to take an account of the change in $g$ and $g_D$ for the different structure sizes. For the following calculations, we use the average value of $g$ and $g_D$ for each structure sizes.

\subsection{Fitting of the experimental data using the SEL}
To find the fracture properties $G_f$ and $c_f$, we combine the experimental results and the stochastic finite element model using Eq.~\eqref{eq6}. To do so, we introduce the following variables:
\begin{equation}\label{eq8}
X = \frac{g}{g_D}D, \;\; Y = \frac{1}{g_D\sigma_{Nc}^{2}}
\end{equation}
Using theses variables, Eq.~\eqref{eq6} can be transformed into the following linear expression:
\begin{equation}\label{eq9}
Y = C + AX
\end{equation}
with:
\begin{equation}\label{eq10}
C = \frac{c_f}{E^*G_f}, \;\; A = \frac{1}{E^*G_f} 
\end{equation}
Finally, the fracture energy, $G_f$, and the effective size of FPZ, $c_f$, are:
\begin{equation}\label{eq11}
G_f = \frac{1}{E^*A}, \;\; c_f = \frac{C}{A} 
\end{equation}

Figures~\ref{f-LR}a-d plot the linear regression analysis introduced in Eqs.~\eqref{eq8} and \eqref{eq9} for the tests conducted in this work from which $G_f$ and $c_f$ can be calculated leveraging Eqs.~\eqref{eq11}. The standard deviations are calculated following \cite{Baz1998}. The measured $G_f$ and $c_f$ for various thicknesses are reported in Table~\ref{T4}. Also, they are plotted in Fig.~\ref{f-Gf} along with a typical Al5083 \cite{Nikos1987} and T700G Quasi-Isotropic (QI) laminate composite \cite{KoArxiv} with a thickness of $3.1$ mm.

As can be noticed, the intra-laminar mode I fracture energy of DFCs are outstanding compared to Al5083, $G_f^{Al} = 12.0$ N/mm and even to the QI laminate composite, $G_f^{QI} = 41.01$ N/mm. In fact, the measured $G_f$ for the DFC plates investigated in this work are $49.54$, $63.47$, $40.32$, and $34.55$ N/mm for thickness of $4.1$, $3.3$, $2.2$, and $1.1$ mm respectively. It is worth noting that these measured $G_f$ are $2.9 - 5.3$ times larger than the typical values of an aluminum alloy which is the main competing material of DFCs in the current aerospace industry. If ones consider the weight saved by using fiber composite structures, having such a high $G_f$ opens new avenues for DFC structures especially for crashworthiness applications. Even comparing with the QI laminate composite made of the identical prepreg system, DFCs with equal thickness provide as much as $1.5$ times higher $G_f$ in the case of $t = $3.3 mm, thanks to the additional energy absorption mechanisms provided by complex mesostructure. 

From Fig.~\ref{f-Gf}, we also observe that $G_f$ depends strongly on the thickness of DFC plates. According to Fig.~\ref{f-Gf}, $G_f$ increases gradually up to about $45.03$ N/mm for a thickness of $\sim 2$ mm. Then, it slowly approaches an asymptotic value of about $57.77$ N/mm for larger thicknesses. As can be noted, the fracture energy seems to follow an exponential trend, $Y=A(1-e^{BX})$ with $A = 57.77$ and $B=-0.76$ although more data points for larger thicknesses are needed to completely clarify this aspect. 

When DFCs contain a sufficient number of platelets, the mesoscale morphology becomes more uniform and the amount of fracture area created in the FPZ tends asymptotically to a limit value, leading to no further changes of $G_f$ with increasing plate thickness. This result provides an important index for the engineers who want to optimize the thickness of DFCs. Increasing the thickness of DFCs will not provide a further benefit in terms of the $G_f$ after $t \sim 3$ mm. For decreasing the thickness, we recommend paying extra attention because the fracture energy starts to drop significantly below $t \sim 2$ mm. The effective FPZ size, $c_f$ also follows the similar trend. The $c_f$ for the thickness of $4.1$, $3.3$, $2.2$, and $1.1$ mm are $3.69$, $8.67$, $3.63$, and $1.78$ mm respectively. This is an agreement with noticeable quasi-ductile fracturing behavior of thicker DFCs.

\subsection{Size Effect Curves}
We can also plot the size effect curves using the experimental values and the finite element modeling results. To do so, the Eq.~\eqref{eq6} is rearranged as follow \cite{Baz1996}:
\begin{equation}\label{eq12}
\sigma_{Nc} = \frac{\sigma_0}{\sqrt{1+D/D_0}} 
\end{equation}
where $\sigma_0 = \sqrt{E^*G_f/c_fg_D}$ and $D_0 = c_fg_D/g$. The normalized size effect curves are plotted using the following axes:
\begin{equation}\label{eq13}
Y' = \frac{\sigma_{Nc}}{\sigma_0}, \;\; X' = \frac{D}{D_0}
\end{equation}

In Fig.~\ref{f-SEL}, the size effect curves are plotted in a double-logarithmic scale. There are two asymptotes in the plots: the horizontal asymptote represents the strength predicted by the strength-based failure criterion, on which the size effect is negligible. The oblique asymptote with a slope of $-1/2$ represents the nominal strength according to LEFM. As can be noticed, regardless of all the explored thicknesses, DFCs show a strong size effect. Additionally, the experimental data clearly show a transition from stress-driven to energy-driven failure. This transition can be explained by considering the relative size of the FPZ compared with the structure size. The relative size of the FPZ, which is a material property, increases as the structure size decreases. Therefore, the nonlinear effect induced by the micro-damage in front of the crack-tip becomes non-negligible. For sufficiently small structures, the effect is so significant to cause a strong deviation from the LEFM. On the other hand, the effect of the FPZ is reduced as the structure size increases and the size effect is well captured by the LEFM.

The characteristic size of the FPZ, $c_f$, increases with the range of thickness tested in this study (see Table~\ref{T4}). The thinnest DFCs are showing the most brittle failure with the smallest effective FPZ size, $c_f$, and fracture energy, $G_f$ (see Fig.~\ref{f-SEL}d). As the thickness increases, DFCs show pronounced pseudo-ductile response (see Fig.~\ref{f-SEL}a). 

Based on the foregoing observations, we can conclude that the strong quasibrittleness must be taken carefully when we design the DFC structures with sharp notches or defects. Neither traditional LEFM nor strength-criteria has the capability to precisely extrapolate the structural strength from the tested specimens, most of which belong to the transitional zone. When estimating the strength of larger structures, LEFM significantly underestimates the strength whereas strength-criteria overestimates it. If we continuously adapt LEFM in designing of DFC structures, the underestimations of strength capability may hinder the possible applications of DFCs in engineering applications. Therefore, a proper model equipped with a characteristic length scale such as SEL or other equivalent models must be utilized to capture the quasibrittleness of DFCs.

\subsection{Brittleness number of DFCs vs traditional laminated composites}
Another useful parameter to compare the fracturing behavior of a structure is called a brittleness number, $\beta$ \cite{Baz1998}. The $\beta$ is defined as the ratio between the structure size $D$ and transition size $D_0$. 
When the $\beta$ is less than 0.1, the structure can be considered as perfectly plastic or quasi-ductile. When the $\beta$ is larger than 10, the structure fractures in a very brittle fashion. If the structure is in-between 0.1 and 10, the size of the FPZ is non-negligible, and the quasibrittleness should be considered. The results are plotted in Fig.~\ref{f-brittle}. Regardless of the specimen sizes, all the experimented thicknesses are well within the quasibrittle zone. As the thickness increases, DFCs quickly deviate from the brittle boundary. The decreasing trend is stronger for the larger specimen sizes. However, after reaching the thickness of $3.3$ mm, we observe a slight increase in the $\beta$. This is in fact the $\beta$ is a function of $c_f$ which estimates the size of the fracture process zone. Irwin \cite{Irwin1958} estimated the size of the inelastic zone in front of the crack tip as $l_{ch} = EG_f/\pi\sigma_{Nc}^2$. Since the both $c_f$ and $l_{ch}$ estimate the size of plastic zone, we can assume that the $c_f$ has a similar function with the $l_{ch}$. As a result, the $\beta$ can be a function of $\sigma_{Nc}^2/EG_f$. Assuming change in $E$ is negligible with respect to the thickness, both $G_f$ and $\sigma_{Nc}$ increase, then saturate at a certain thickness. However, $G_f$ could reach the saturation point earlier than $\sigma_{Nc}$. Such unbalance in saturation thickness can cause the change in slope of $\beta$ with respect to the thickness. After $G_f$ and $\sigma_{Nc}$ both reach the saturation point, the $\beta$ will remain constant. Further investigation is in progress using the computational model to confirm the trend.





\section{Conclusions}
Combining experiments and stochastic finite element modeling, this work studied the fracturing behavior and scaling effect of Discontinuous Fiber Composite (DFC) structures with different thicknesses. The following conclusions are drawn based on the results of this study:

\be  \setlength{\itemsep}{1mm}
\ii The experimental results on geometrically-scaled Single Edge Notch Tension specimens of four distinct thicknesses showed a significant size effect on the nominal strength of DFC structures. For a given thickness, smaller specimens exhibited a pronounced pseudo-ductile fracture behavior with minimum scaling effect. In contrast, when the size of the specimen was sufficiently large, the scaling of the nominal strength approached Linear Elastic Fracture Mechanics (LEFM) asymptotically and fracture occurred in a very brittle manner;

\ii The transition from pseudo-ductile to brittle fracture with an increasing specimen size is related to the development of a significant Fracture Process Zone (FPZ) whose dimensions were found to be comparable to the platelet size. In the FPZ, signiﬁcant non-linear deformations due to sub-critical damage mechanisms, such as platelet delamination, matrix microcracking, and platelet splitting/fracture, promote strain redistribution and mitigate the intensity of the stress field induced by the crack/notch. This phenomenon is more pronounced for small structures since the size of a fully-developed FPZ is typically a material property and thus its influence on the structural behavior becomes increasingly significant as the structure size is reduced. For sufficiently large structures, the size of the FPZ becomes negligible compared to the structure's characteristic size in agreement with the inherent assumption of the LEFM that non-linear effects are negligible during the fracturing process;

\ii A significant effect of the thickness was also found on the nominal strength of DFC structures and its Coefficient of Variation (CoV). For a given specimen size, thinner specimens tended to fracture in a relatively brittle way with the nominal strength being closer to LEFM compared to thicker specimens. At the same time, the scatter of the experimental data increased with decreasing thickness from values in the order of $9.31\%$ for a thickness of $4.1$ mm to $21.26\%$ for a thickness of $1.1$ mm for the size $1$ specimens;

\ii The highest scatter of the tests on thinner specimens was confirmed by the analysis of the fracture morphology. For a given specimen size, the fracture process in thick DFCs was mostly driven by the FPZ developing at the notch tip. In contrast, the fracture of thin specimens was often initiated far from the notch, showing that failure was driven by random weak spots in the material rather than the notch. This phenomenon is related to the average number of platelets through the thickness: when this number is low, the probability of having weak spots with platelets that are not favorably oriented with respect to the load increases if one assumes the distribution mechanism of platelets is not affected by the thickness.
This was clearly showed in this work by numerical simulations that reported a shift of the PDF of average platelet orientation through the thickness towards larger angles compared to the load axis with decreasing thickness;

\ii To investigate the effect of the plate thickness on the fracture energy, $G_f$, and the effective length of the fracture process zone, $c_f$, the approach combining equivalent fracture mechanics and stochastic finite element modeling proposed in \cite{KoArxiv} was used. This model accounts for the effects of the complex random mesostructure of the material by modeling the platelets explicitly. This theoretical framework was able to describe the scaling of structural strength and enabled the characterization of the mode I fracture energy of DFCs;

\ii $G_f$ and $c_f$ were estimated for a platelet size of $50\times8$ mm, and a structure thickness of $4.1$, $3.3$ $2.2$, and $1.1$ mm respectively. It was found that $G_f=$ $49.54\pm9.57$ N/mm, $63.47\pm14.16$ N/mm, $40.32\pm14.98$ N/mm, and $34.55\pm9.50$ N/mm while $c_f=$ $3.69\pm0.51$ mm, $8.67\pm1.37$ mm, $3.63\pm0.97$ mm, and $1.78\pm0.30$ mm. These results clearly indicate a strong effect of the thickness on the fracture properties of the material. In particular, the fracture energy was found to increase gradually with the plate thickness up to an asymptotic value of about $58$ N/mm when the thickness becomes larger than $3$ mm. Further computational studies are ongoing to confirm this trend and to extend the study to other platelet sizes and plate thicknesses;

\ii For all the thicknesses investigated in this work, the analysis of the fracture tests highlighted outstanding fracture energy of DFCs, from $2.9$ to $5.3$ times larger than the one of a typical Al5083 or a Quasi-Isotropic laminate made from the same prepregs for the platelet size investigated in this work. This result is particularly interesting in view of a possible use of DFCs for crashworthiness applications;


 \ee
 
 The critical investigation of the foregoing results can pave the way for the development of novel strategies for the tuning of the fracturing behavior leveraging the DFC mesostructural morphology.


\section*{Acknowledgments}

This study is financially supported by the FAA-funded Center of Excellence for Advanced Materials in Transport Aircraft Structures (AMTAS) and the Boeing Company. Partial support is also provided by the Joint Center for Aerospace Technology Innovation (JCATI). We thank Ahmet Oztekin, Cindy Ashforth, and Larry Ilcewicz from the FAA, and William Avery from the Boeing Company for their guidance and support. We also thank the technical support provided by Bruno Boursier from the Hexcel Corporation.


\begin{appendices}
\section{DFC mesostructure generation algorithm}
\label{AppxA}

The present algorithm is an extension of the stochastic laminate analogy proposed in \cite{Fera2010, Tuttle2017, Sel2017}. The structure of interest is first divided into about $1$ mm by $1$ mm partitions. This partition size is chosen based on a balance between accuracy and computational cost for the given platelet size \cite{KoArxiv}. For a different platelet size, the partition size should be recalculated. A single platelet with predefined length and width is generated over the structure (see Fig.~\ref{f-FEMGeo}a). The platelet center point and the orientation are assigned following the uniform probability \cite{KoArxiv}. For more complex structures, the manufacturing process should be simulated explicitly to find the platelet orientation distribution accurately \cite{Favaloro2018}. Alternatively, the platelet morphology can be characterized by nondestructive evaluation techniques such as micro-computed tomography \cite{Wan2018, Denos2018a}. The platelet information (e.g. platelet orientation) is then saved for each partition within the boundary of the platelet.

Besides simulating the random in-plane distributions of the platelets, it is also important to capture the local thickness variations. For DFCs, the local thickness is a spatial random variable. We found that the CoV for the average platelets through the thickness ($\sim24$) was $0.22$ \cite{KoArxiv}. This CoV is applied for all the other thickness plates because it corresponds to the identical manufacturing process, platelet size, and prepreg material. 

To match the foregoing morphological information, we control the platelet generation following specific rules. First, a platelet-limit zone is created to limit the deposition of the platelets at certain partitions. If the number of platelets in the partitions reaches $\mu\left(1+\text{CoV}\right)$ with $\mu=$ desired average number of platelets, the limit zone is assigned to reject further depositions. Second, the platelet generation process is subdivided into multiple stages called saturation steps. At each saturation step, the CoV is enforced on the current system of platelets. An example of a completed mesostructure with the target average platelets through thickness equals to $24$ is shown in Fig.~\ref{f-FEMGeo}b whereas Fig.~\ref{f-FlowChart} outlines the logic of the platelet generation algorithm.

After the mesostructure generation is completed, the thickness of the individual partitions is adjusted to have a uniform total thickness. A uniform thickness is considered since the CoV on the thickness of the specimens is between $4 - 7\%$. The reason for such a low CoV in measured thickness is the resin flow. Since the present model does not explicitly model the flow, the thickness adjustment of individual partitions is necessary. Two scenarios are considered: (1) if the partition thickness is larger than the mean value, the thickness of the individual platelets is reduced linearly to match the mean thickness. (2) if the partition thickness is lower than the mean value, the thickness of individual platelets does not change. Instead, layers of resin having the matrix elastic constituents are introduced in-between the platelets. 

\section{Computation of $g(\alpha)$ and $g_D(\alpha)$ and the fracture energy}
\label{AppxB}
\renewcommand{\theequation}{B.\arabic{equation}}
\setcounter{equation}{0}

In this work the definition of $G$ is used directly to calculated the energy release rate functions \cite{Baz1998}:
\begin{equation}\label{eq7}
G(\delta,a) = -\frac{1}{t}\left[\frac{\partial\Pi(\delta,a)}{\partial a}\right]_\delta
\end{equation}
where $\delta$ is the equilibrium displacement, $t$ is the thickness, and $\Pi$ is the potential energy of the structure. 

In the Abaqus/Standard \cite{Abaqus} model, the platelets and the resins are assumed to be linear elastic with the material properties listed in Table~\ref{T3}. A mesh of $8$-node, quadrilateral Belytschko-Tsay shell elements with reduced integration (S8R) is used with the maximum size of the element equal to the size of the partitions. A uni-axial uniform displacement is applied at the end where the other end is fixed in all directions. The reaction force, $P$ as well as the $\Pi$ are recorded.

$10$ different lengths of cracks are simulated for each structure. The size of a crack increment is equal to the size of the partition divided by $20$. This results in $\Delta\alpha \approx 0.008$ for $D = 6.5$ mm. In a previous study \cite{Sal2016a}, $\Delta\alpha = 0.0025$ was used. Then, $g$ and $g_D$ are calculated from Eq.~\eqref{eq2}. Figure~\ref{f-findg} shows a typical plot of $g$ against the normalized crack length, $\alpha$. The linear regression analysis is performed to find the slope, $g_D$ at $\alpha=\alpha_0$. 

\end{appendices}



\clearpage

\begin{figure}
 \centering
  \includegraphics[trim=0cm 0cm 0cm 0cm, clip=true,width = 0.4\textwidth, scale=1]{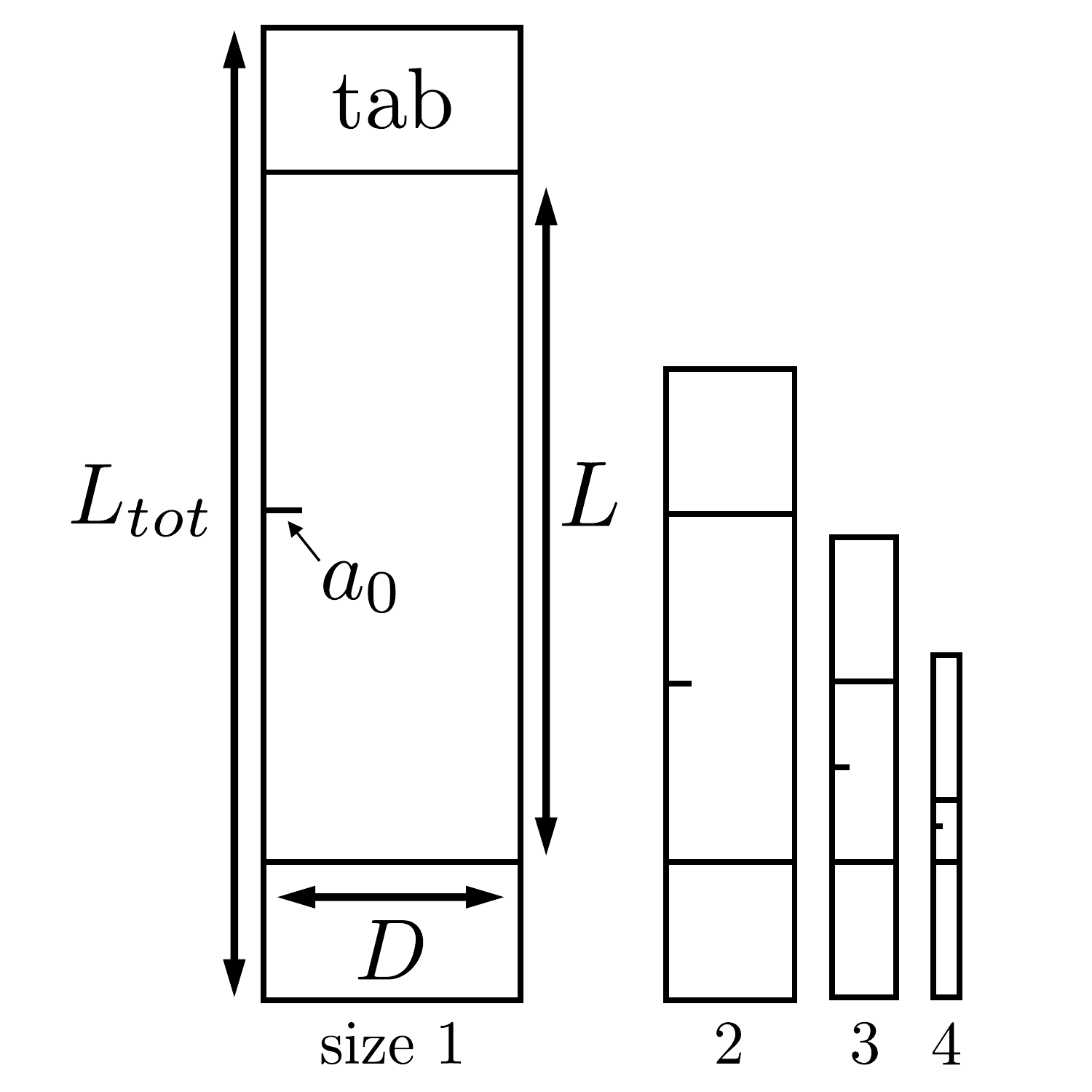} \caption{\label{f-Geometry} \sf Geometry details of the Single Edge Notch Tension specimens.} 
\end{figure}

\begin{figure}
 \centering
  \includegraphics[trim=0cm 0cm 0cm 0cm, clip=true,width = 0.8\textwidth, scale=1]{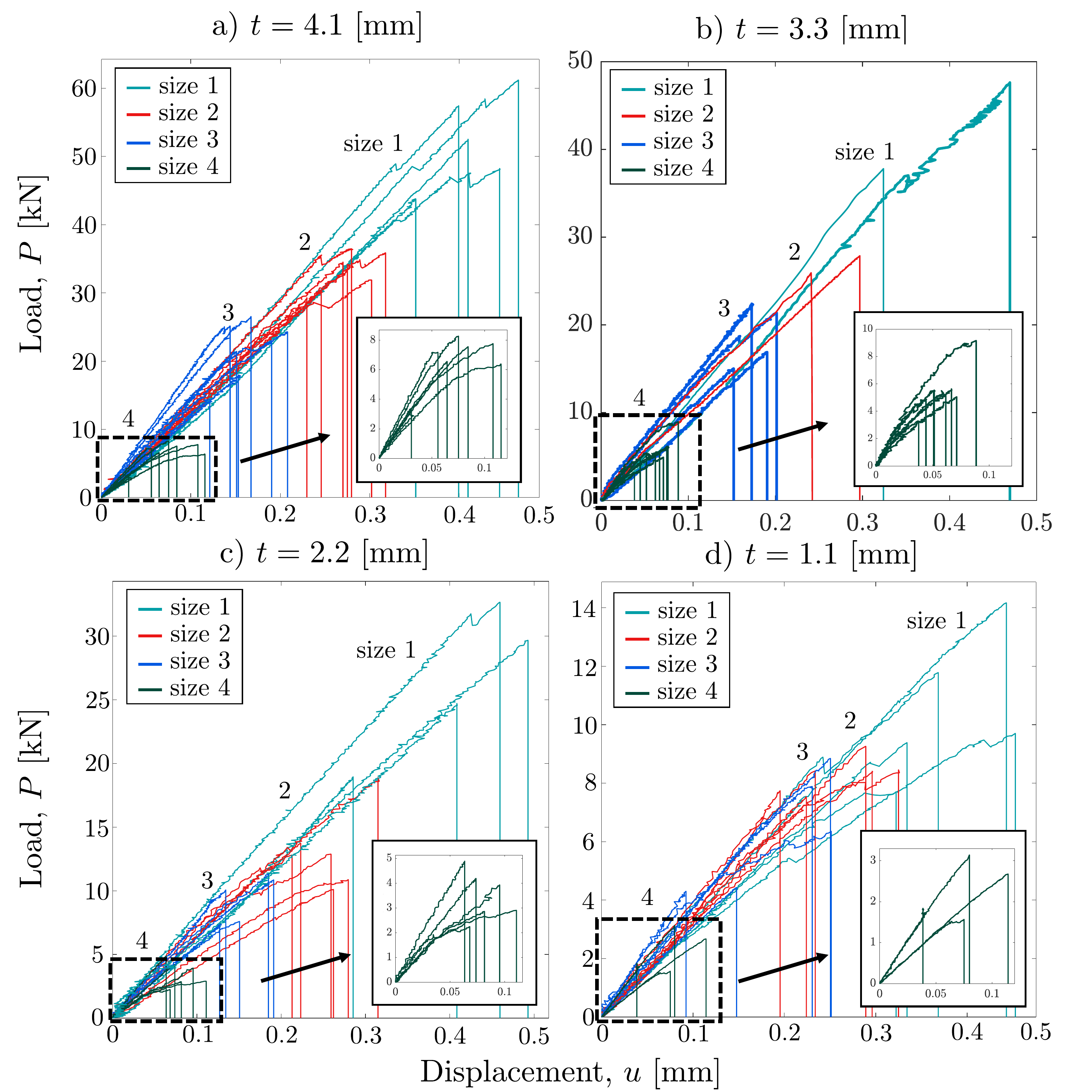} \caption{\label{f-FDdiagram} \sf Load-displacement curves of DFCs with the thickness of a) $4.1$ mm, b) $3.3$ mm, c) $2.2$ mm, and d) $1.1$ mm.} 
\end{figure}

\begin{figure}
 \centering
  \includegraphics[trim=0cm 0cm 0cm 0cm, clip=true,width = 0.8\textwidth, scale=1]{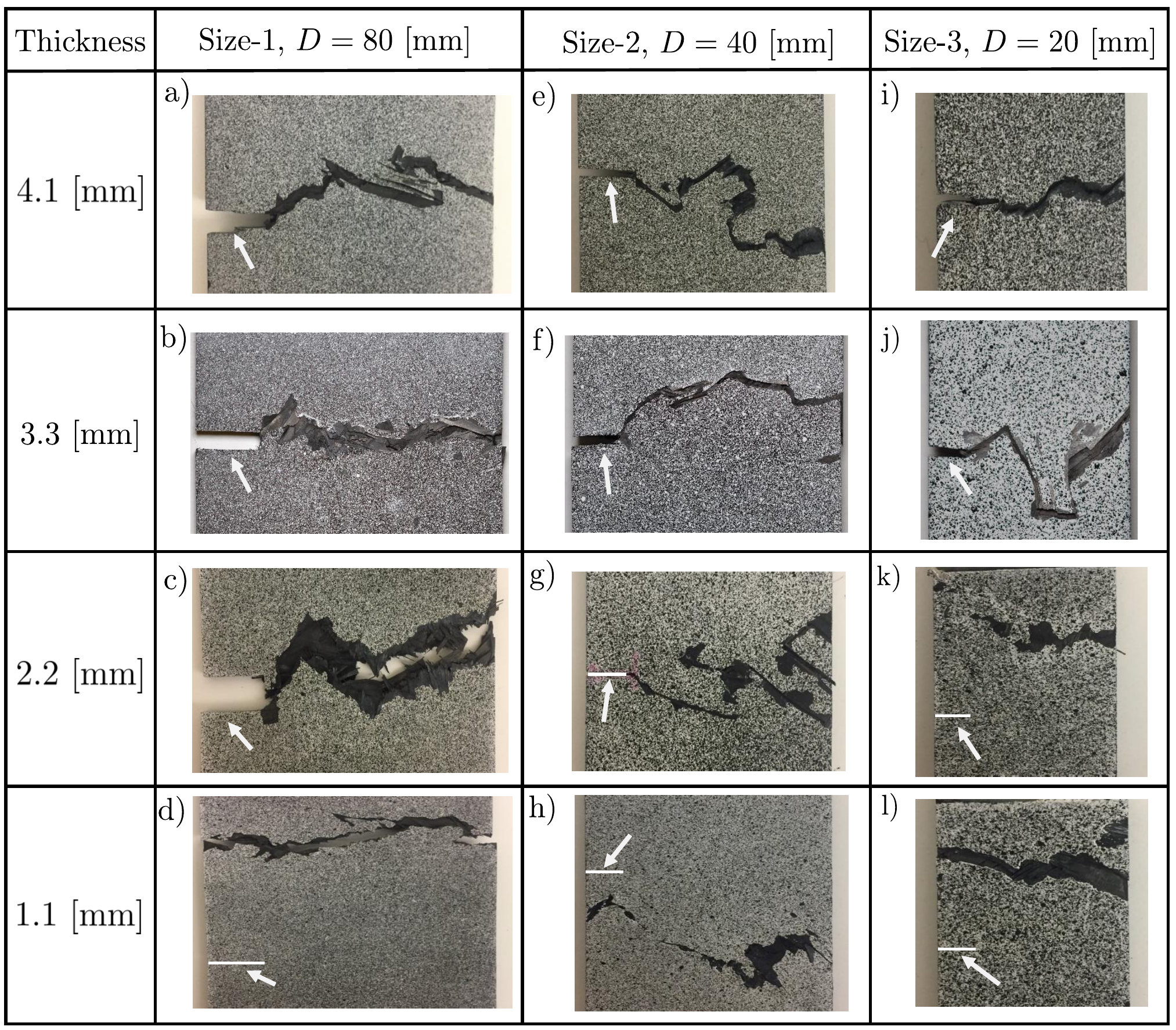} \caption{\label{f-FractureSurfaces} \sf Representative fracture surfaces of Single Edge Notch Tension specimens for all experimental cases. An arrow indicates an initial notch position.} 
\end{figure}

\begin{figure}
 \centering
  \includegraphics[trim=0cm 0cm 0cm 0cm, clip=true,width = 1\textwidth, scale=1]{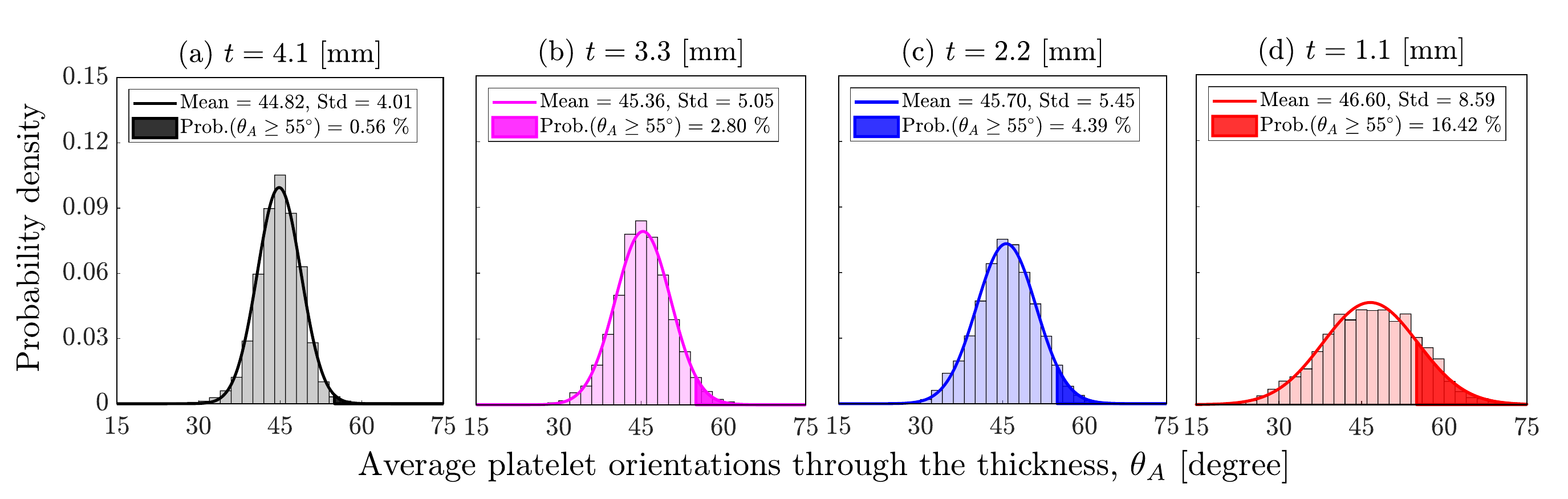} \caption{\label{f-pdf} \sf Probability density distributions of the average platelet orientations through the thickness, $\theta_A$, obtained by simulation. Plate thickness: (a) $4.1$ mm, (b) $3.3$ mm, (c) $2.2$ mm, and (d) $1.1$ mm.}
\end{figure}

\begin{figure}
 \centering
  \includegraphics[trim=0cm 1cm 0cm 1cm, clip=true,width = 0.8\textwidth, scale=1]{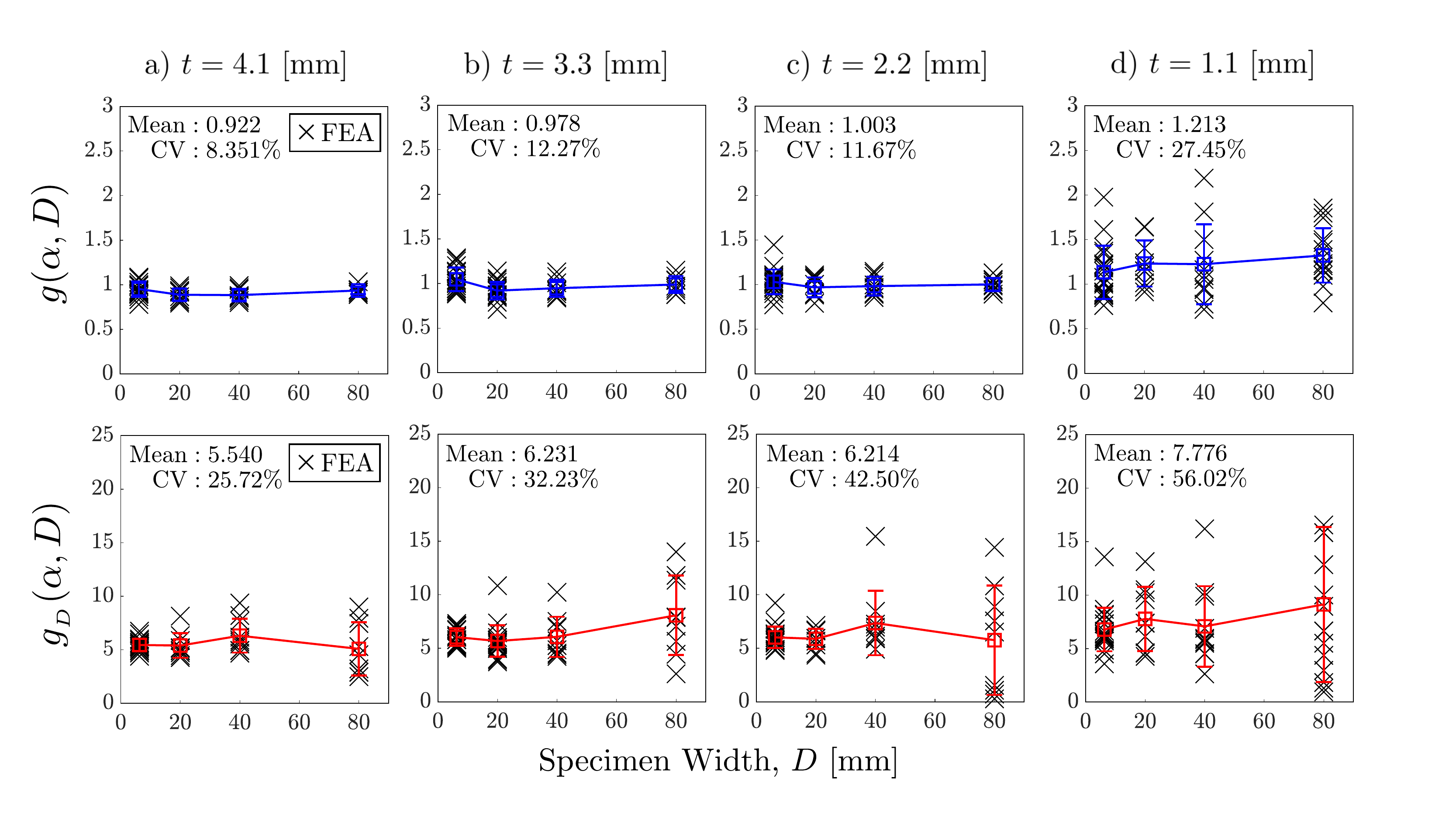} \caption{\label{f-gandgprime} \sf Dimensionless energy release rate parameters, $g$ and $g_D$ for the thickness of (a) $4.1$ mm, (b) $3.3$ mm, (c) $2.2$ mm, and (d) $1.1$ mm.} 
\end{figure}

\begin{figure}
 \centering
  \includegraphics[trim=0cm 0cm 0cm 0cm, clip=true,width = 0.7\textwidth, scale=1]{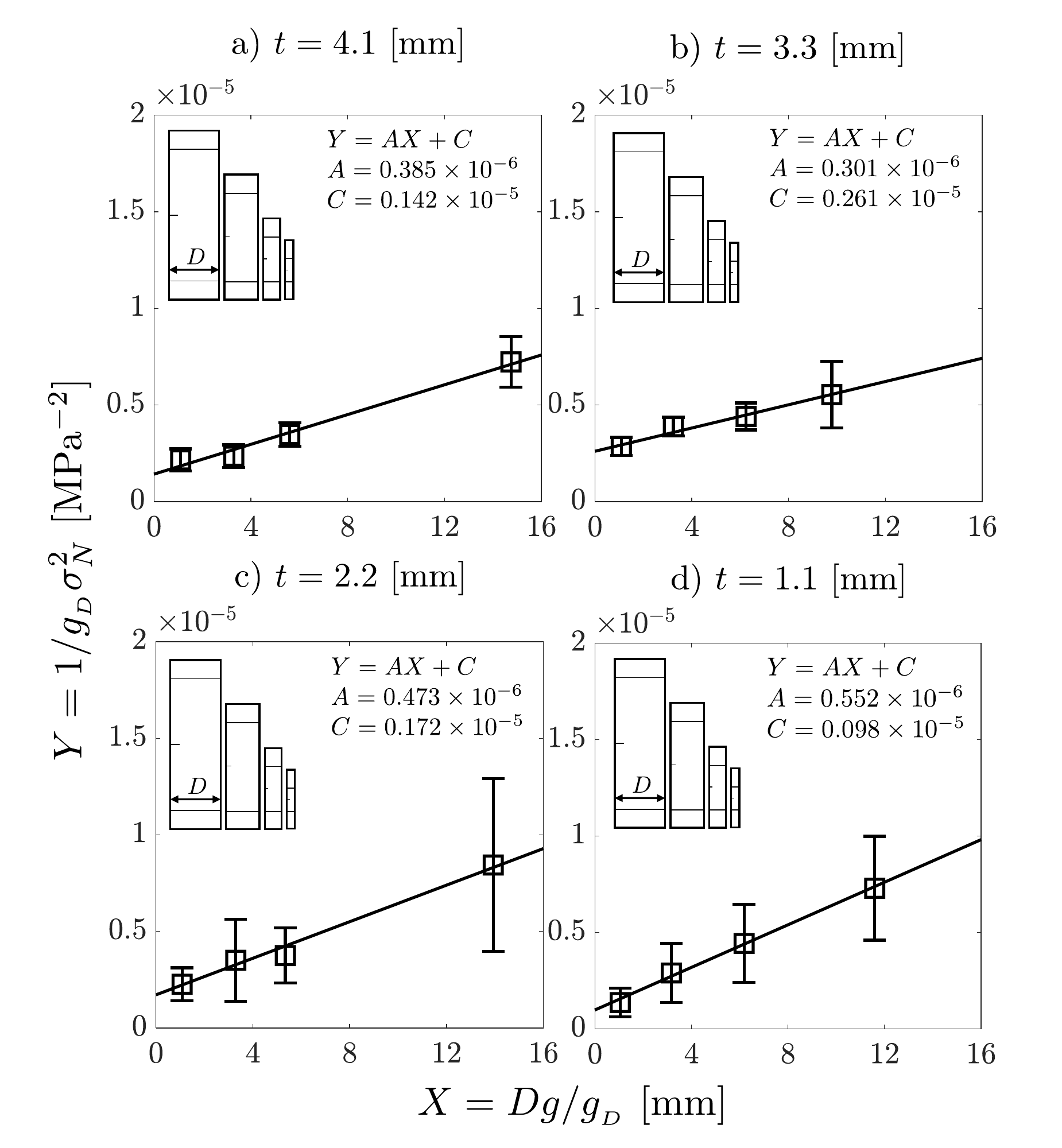} \caption{\label{f-LR} \sf Linear regression analysis to find the fracture properties of DFCs with thickness of (a) $4.1$ mm, (b) $3.3$ mm, (c) $2.2$ mm, and (d) $1.1$ mm.}
\end{figure}

\begin{figure}
 \centering
  \includegraphics[trim=0cm 0cm 0cm 0cm, clip=true,width = 0.5\textwidth, scale=1]{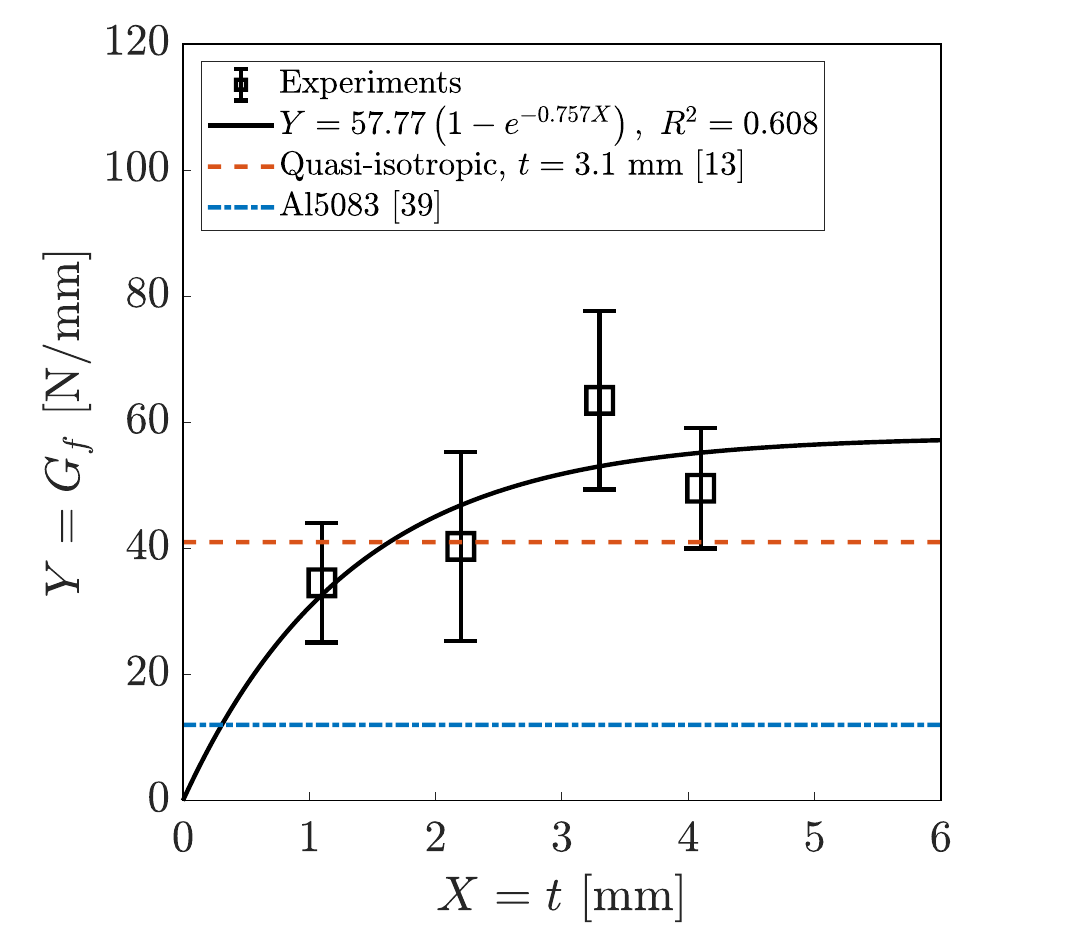} \caption{\label{f-Gf} \sf Measured fracture energy of DFCs with the exponential least square fitting.}
\end{figure}

\begin{figure}
 \centering
  \includegraphics[trim=0cm 0cm 0cm 0cm, clip=true,width = 1\textwidth, scale=1]{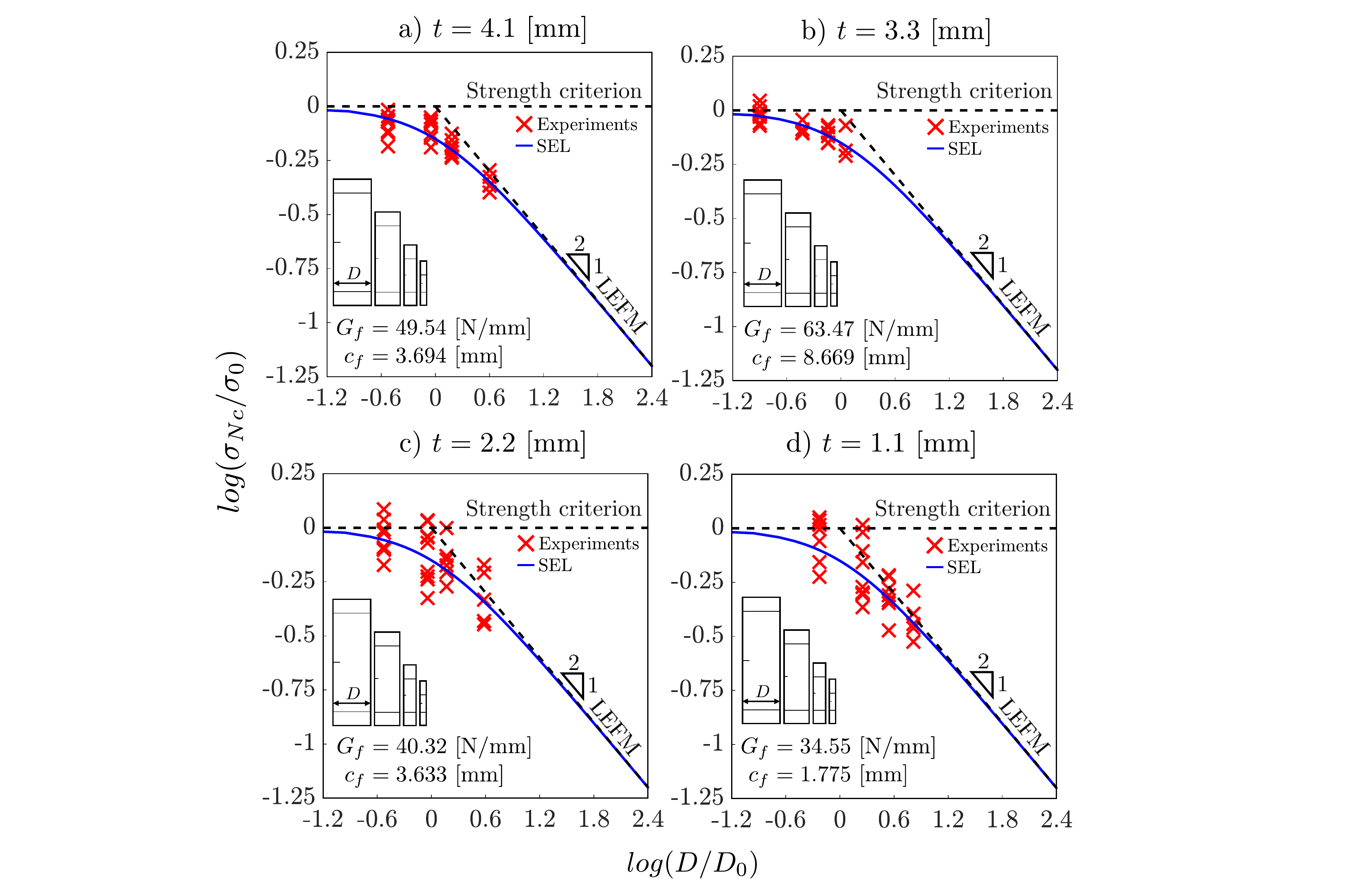} \caption{\label{f-SEL} \sf Measured size effects for DFCs with thickness of (a) $4.1$ mm, (b) $3.3$ mm, (c) $2.2$ mm, and (d) $1.1$ mm.}
\end{figure}

\begin{figure}
 \centering
  \includegraphics[trim=0cm 0cm 0cm 0cm, clip=true,width = 0.8\textwidth, scale=1]{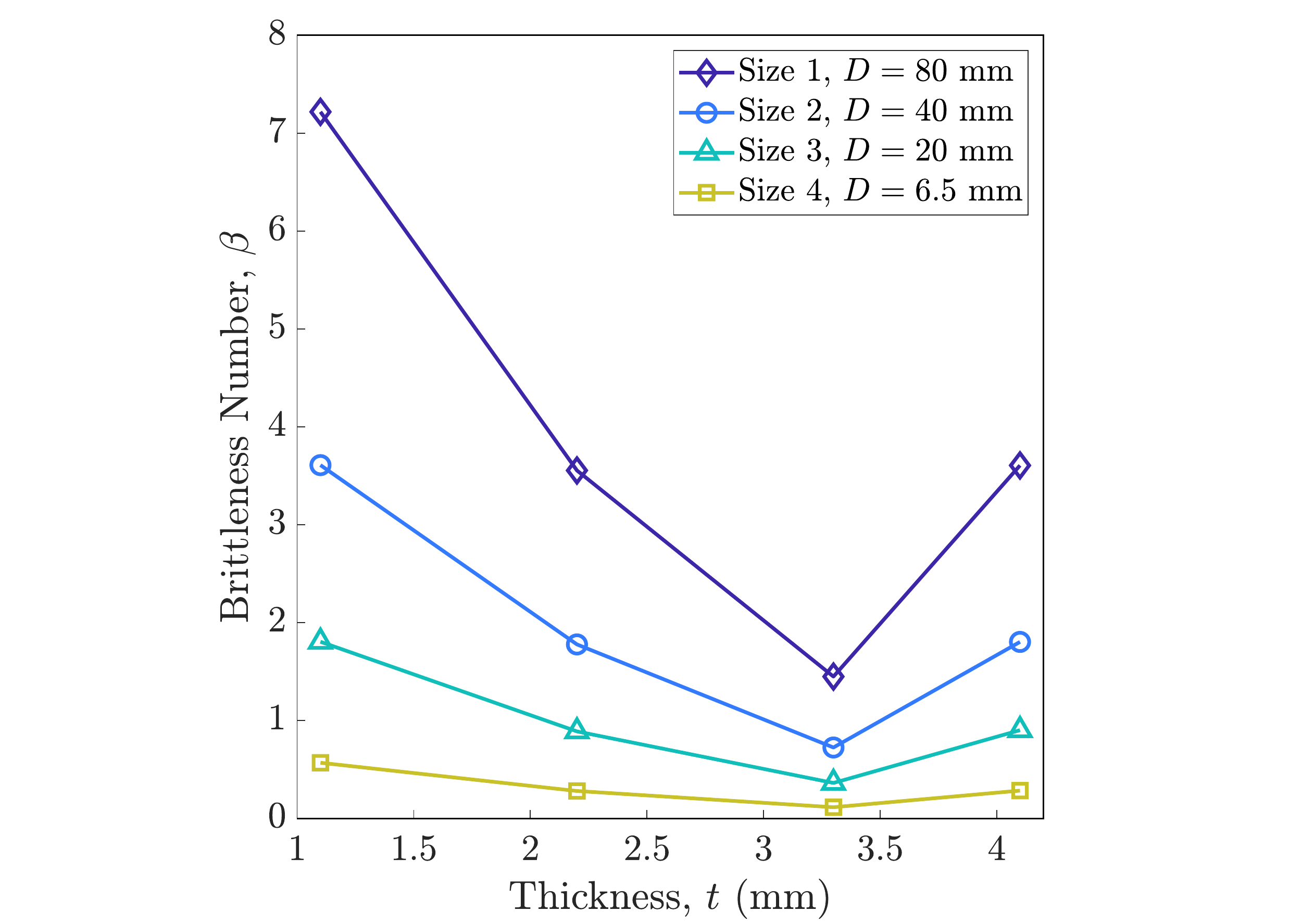} \caption{\label{f-brittle} \sf Change of the brittleness number, $\beta$, as a function of the thickness for all the experimented sizes.}
\end{figure}

\renewcommand{\thefigure}{A\arabic{figure}}
\setcounter{figure}{0}

\begin{figure}
 \centering
  \includegraphics[trim=0cm 0cm 0cm 0cm, clip=true,width = 1\textwidth, scale=1]{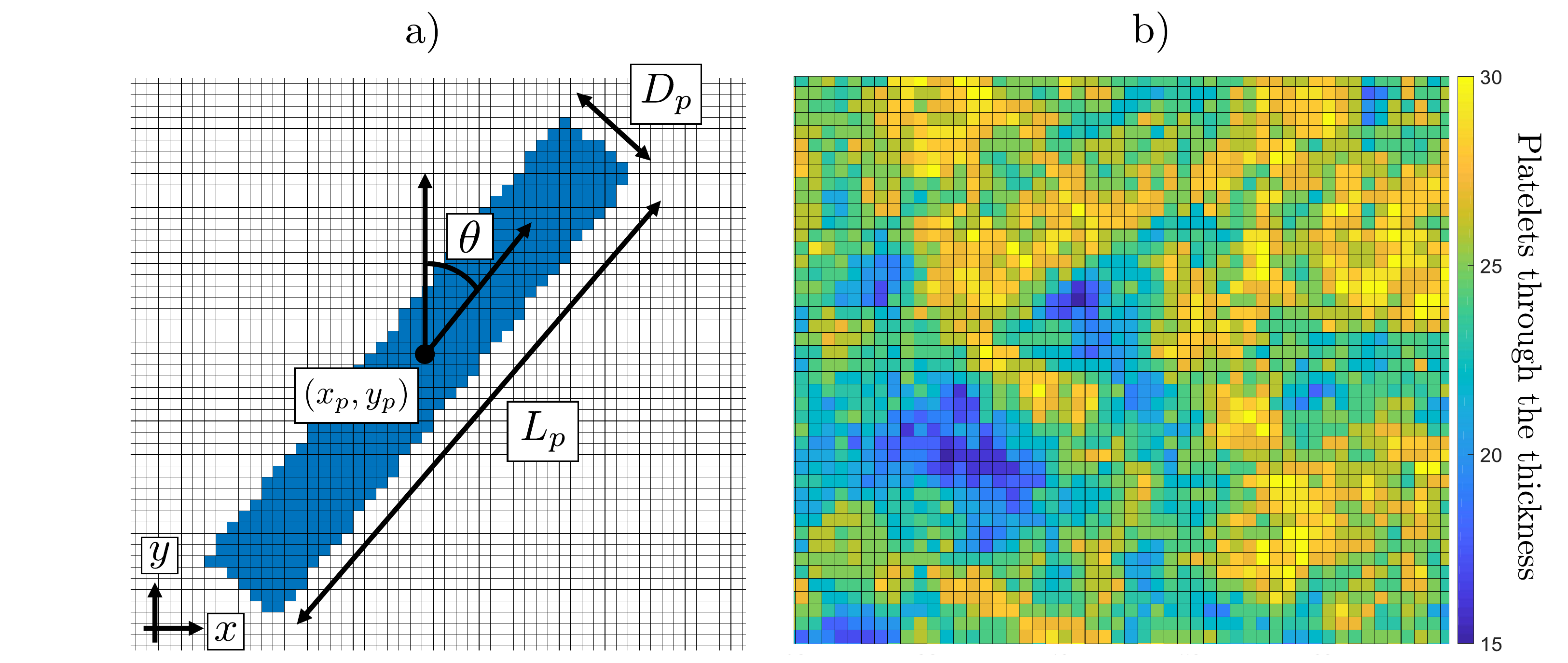} \caption{\label{f-FEMGeo} \sf (a) A platelet geometry in partitions. The platelet's length and width ($L_p, D_p$) are $50\times8$ mm. The mesostructure algorithm chooses the center point ($x_p, y_p$) and orientation ($\theta$) based on the uniform distribution.  (b) A sample $50\times50$ mm plate with an average of $24$ platelets through the thickness. } 
\end{figure}

\begin{figure}
 \centering
  \includegraphics[trim=0cm 0cm 0cm 0cm, clip=true,width = 1\textwidth, scale=1]{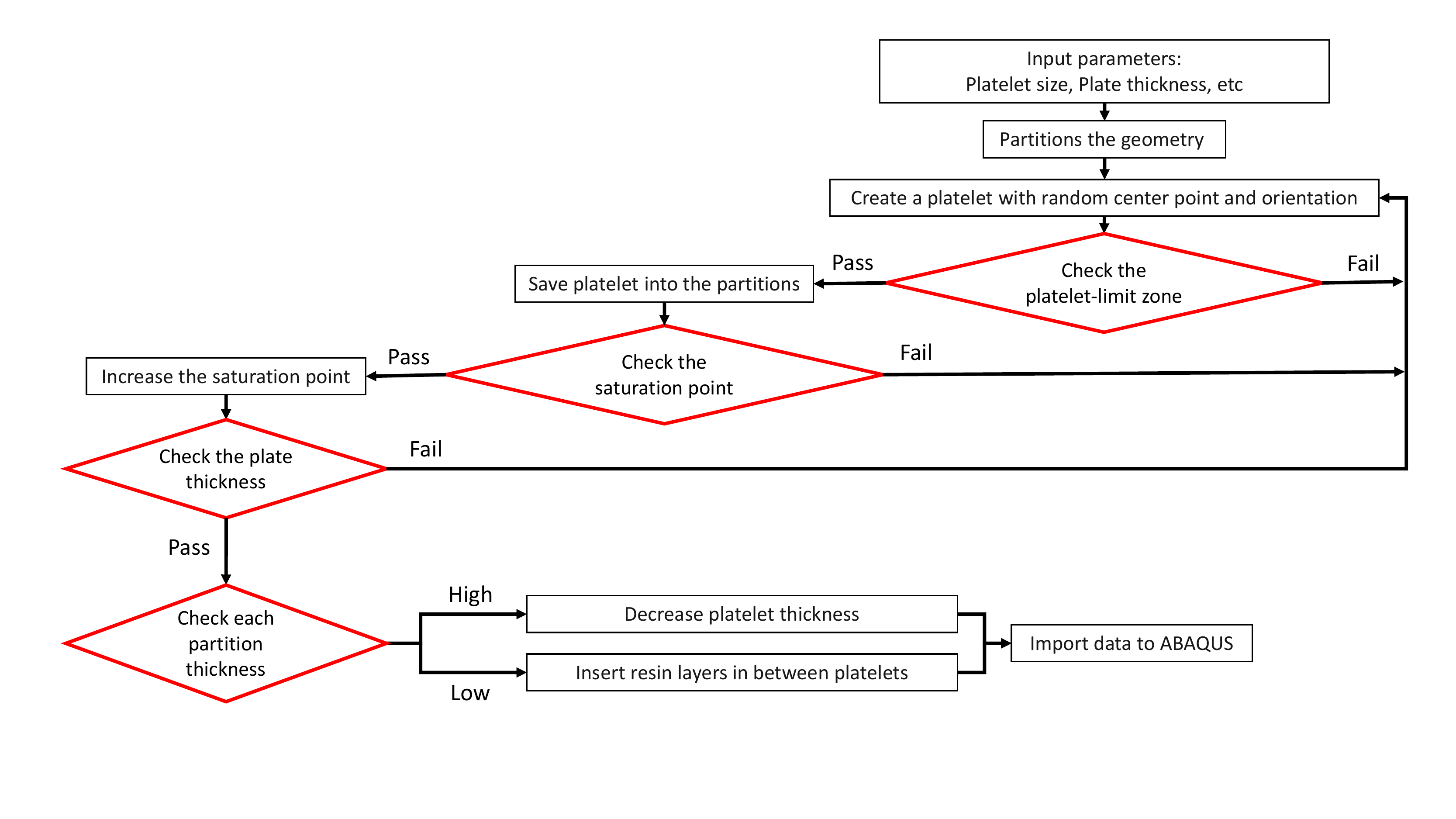} \caption{\label{f-FlowChart} \sf The DFC mesostructure generation algorithm flow chart.} 
\end{figure}

\renewcommand{\thefigure}{B\arabic{figure}}
\setcounter{figure}{0}

\begin{figure}
 \centering
  \includegraphics[trim=0cm 0cm 0cm 0cm, clip=true,width = 0.6\textwidth, scale=1]{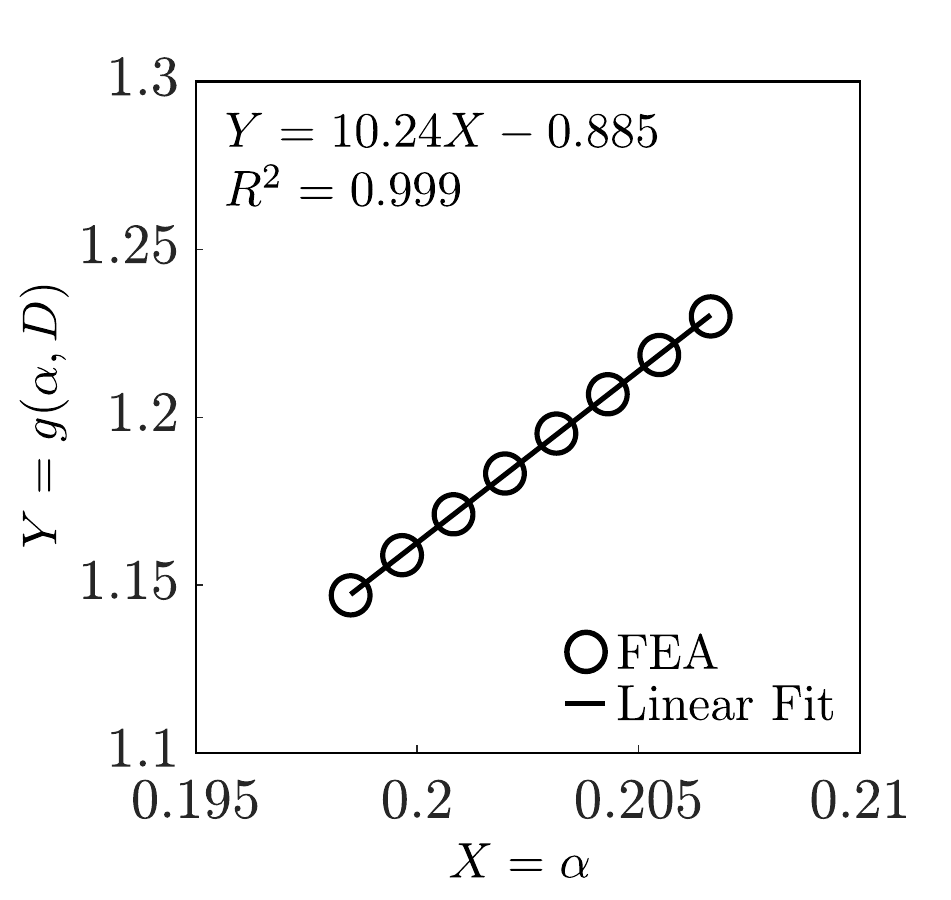} \caption{\label{f-findg} \sf A sample calculation of dimensionless energy release rate parameters, $g$ and $g_D$.} 
\end{figure}

\clearpage

\begin{table}[ht]
\centering  
\caption{\sf Geometry information of the Single Edge Notch Tension (SENT) specimens.}
\begin{adjustbox}{max width=\textwidth}
\begin{tabular}{ c c c c c } 
 \hline
  \rule{0pt}{4ex}
 Size & Width, $D$  & Gauge length, $L$  & Total length, $L_{Total}$  & Notch length, $a_0$ \\
  & (mm) & (mm) & (mm) & (mm)\\[1 ex]
 \hline
$1$ & $80$ & $178$ & $254$ & $16$ \\
$2$ & $40$ & $89$ & $165$ & $8$ \\
$3$ & $20$ & $44.5$ & $120.5$ & $4$ \\
$4$ & $6.5$ & $14.5$ & $90.5$ & $1.3$ \\
\hline
\multicolumn{5}{l}{$^{\mbox{*}}$ Tested thicknesses are $4.1$, $3.3$, $2.2$, $1.1$ mm with the platelet size of $50 \times 8$ mm.}
\end{tabular}
\end{adjustbox}
\label{T1}
\end{table}

\begin{table}[ht]
\centering  
\caption{\sf The average failure strength with standard deviation of tested SENT specimens.}

\begin{adjustbox}{max width=\textwidth}
\begin{tabular}{c c c c c}
\hline
& \multicolumn{4}{c}{Failure strength, $\sigma_{Nc}$ (MPa)} \\
       \cline{2-5}
Thickness (mm) & Size1   & Size2   & Size3   & Size4  \\
\hline
$4.1$ & $166.7 \pm 15.53$ & $216.0 \pm 20.31$ & $285.8 \pm 32.32$  & $297.7 \pm 35.27$ \\
$3.3$ & $153.8 \pm 27.61$ & $198.2 \pm 15.54$ & $214.2 \pm 14.58$  & $242.2 \pm 20.14$  \\
$2.2$ & $158.2 \pm 45.90$ & $200.5 \pm 44.11$ & $248.0 \pm 71.04$  & $284.3 \pm 55.82$  \\
$1.1$ & $128.5 \pm 27.32$ & $180.0 \pm 33.82$ & $239.0 \pm 80.28$  & $354.8 \pm 78.27$  \\
\hline
\end{tabular}
\end{adjustbox}
\label{T2}
\end{table}

\begin{table}[ht]
\centering  
\caption{\sf The fracture properties and dimensionless energy release rate parameters calculated from the experiments and the stochastic FEM.}

\begin{adjustbox}{max width=\textwidth}
\begin{tabular}{ c c c c c} 
 \hline
  \rule{0pt}{4ex}
 Thickness & Fracture energy, $G_f$  & Effective FPZ length, $c_f$ & $g(\alpha_0)^*$  & $g_D(\alpha_0)^*$\\
 (mm) & (N/mm) & (mm) & - & -\\[1 ex]
 \hline
$4.1$ & $49.54 \pm 9.57$ & $3.69 \pm 0.51$ & $0.92 \pm 0.08$ & $5.54 \pm 1.43$ \\
$3.3$ & $63.47 \pm 14.16$ & $8.67 \pm 1.37$ & $0.98 \pm 0.12$ & $6.23 \pm 2.01$ \\
$2.2$ & $40.32 \pm 14.98$ & $3.63 \pm 0.97$ & $1.00 \pm 0.12$ & $6.21 \pm 2.64$ \\
$1.1$ & $34.55 \pm 9.50$ & $1.78 \pm 0.30$ & $1.21 \pm 0.33$ & $7.78 \pm 4.36$ \\

\hline
\multicolumn{5}{l}{${\mbox{*}}$ Averaged FE simulation results of each thicknesses.}
\end{tabular}
\end{adjustbox}
\label{T4}
\end{table}

\renewcommand{\thetable}{A\arabic{table}}
\setcounter{table}{0}

\begin{table}[ht]
\centering  
\caption{\sf Elastic material properties for the T$700$G and matrix layers.}

\begin{adjustbox}{max width=\textwidth}
\begin{tabular}{ c c c} 
 \hline
  \rule{0pt}{4ex}
 Properties &  T$700$G  & Matrix  \\ [1 ex]
 \hline
Platelet initial thickness, $t_p$ [mm] & $0.139$ & Varies\\
Longitudinal modulus, $E_{1}$ [GPa] & $135$ & $3$ \\
Transverse modulus, $E_{2}$ [GPa] & $10$ & $3$ \\
Shear modulus, $G_{12}$ [GPa] & $5$ & $1.1$\\
Poisson ratio, $\nu_{12}$ & $0.3$ & $0.35$ \\

\hline
\end{tabular}
\end{adjustbox}
\label{T3}
\end{table}

\clearpage


\end{document}